\documentclass[twocolumn]{aastex631}

\usepackage{amsmath}
\usepackage{amssymb}
\usepackage{mathrsfs}
\usepackage[T1]{fontenc}
\usepackage{natbib}

\renewcommand{\edit}[2]{#2}

\begin{document}

\title{Application of the disk instability model to all Quasi-Periodic Eruptions}

\author[0000-0002-6938-3594]{Xin Pan}
\affiliation{Key Laboratory for Research in Galaxies and Cosmology, Shanghai Astronomical Observatory, Chinese Academy of Sciences, 80 Nandan Road, Shanghai 200030, People's Republic of China}
\affiliation{University of Chinese Academy of Sciences, 19A Yuquan Road, 100049, Beijing, People's Republic of China}

\author[0000-0002-7299-4513]{Shuang-Liang Li}
\affiliation{Key Laboratory for Research in Galaxies and Cosmology, Shanghai Astronomical Observatory, Chinese Academy of Sciences, 80 Nandan Road, Shanghai 200030, People's Republic of China}

\author[0000-0002-2355-3498]{Xinwu Cao}
\affiliation{Institute for Astronomy, Zhejiang Institute of Modern Physics, School of Physics, Zhejiang University, 866 Yuhangtang Rd, Hangzhou, 310058, People's Republic of China}

\correspondingauthor{Xin Pan, Shuang-Liang Li, Xinwu Cao}
\email{panxin@shao.ac.cn, lisl@shao.ac.cn, xwcao@zju.edu.cn}


\begin{abstract}

After the first quasi-periodic eruptions (QPEs, GSN069) was reported in 2019, four other sources
have been identified as QPEs or its candidate. However, the physics behind QPEs is still unclear so far, though several models have been proposed. \cite{2022ApJ...928L..18P} proposed an instability model for the accretion disk with magnetically driven outflows in the first QPEs GSN 069, which is able to reproduce both the light curve and the evolution of spectrum fairly well. In this work, we exploit this model to all the QPEs. We imporve the calculations of the spectrum of disk by introducing a hardening factor, which is caused by the deviation of opacity from the blackbody. 
 We find that the light curves and evolution of the spectra of the four QPEs or candidate can all be well reproduced by our model calculations.
\end{abstract}

\section{Introduction}

After the first discovery of Quasi-periodic eruptions (QPEs) in GSN 069, four other QPEs sources, i.e., RX J1301.9+2747, eRO-QPE1, eRO-QPE2 and XMMSL1 J024916.6-041244 (most possible) have been discovered \citep{2019Natur.573..381M,2020A&A...636L...2G,2021Natur.592..704A,2021ApJ...921L..40C}. All the QPEs show some similar features, such as the short eruption periods, high-amplitude bursts, and occurring mainly in the soft X-ray band. The primary challenges of this phenomenon are how to construct a physical scenario to produce such a shortly periodic eruptions (several to a dozen of hours).
A number of models have been proposed, which can be roughly divided into two categories: while the first one suggests that the periodic outbursts in QPEs originates from the periodic orbital motion of a star captured by the black hole \citep{2020MNRAS.493L.120K,2021MNRAS.503.1703I,2021ApJ...921L..32X,2021ApJ...917...43S,2022ApJ...926..101M,2022ApJ...933..225W,2022arXiv220902786K,2022MNRAS.515.4344K,2022arXiv221008023L,2022arXiv221009945C,2022arXiv221109851L}, another one ascribes the periodic behavior to the instability of inner accretion disk dominated by the radiation pressure \citep{2020A&A...641A.167S,2021ApJ...910...97P,2021ApJ...909...82R,2022ApJ...928L..18P,2022arXiv220410067S,2022arXiv221100704K}. Notably, only the model of \citet{2022ApJ...928L..18P} is able to fit both the light curves and the phase-resolved X-ray spectrum simultaneously during outbursts in GSN 069. 

It was suggested there is some evidence of tidal disruption events (TDE) in two QPEs (GSN 069 and XMMSL1 J024916.6-041244) \citep{2007A&A...462L..49E,2018ApJ...857L..16S,2021ApJ...920L..25S}. If this is the case, a remnant of core or a white dwarf that continually rotates around the black hole may produce QPEs by partial TDEs \citep{2021ApJ...920L..25S,2023A&A...670A..93M}. However, such a model may only apply for these two QPEs, and the detailed physical processes of the such TDE evolution and their radiation properties are still quite unclear \citep{2015ApJ...806..164P,2016MNRAS.461..948M,2021SSRv..217...16B}. 


It is well known that the inner part of a thin disk dominated by radiation pressure is both thermally and viscously unstable leading to limit-cycle behaviours \citep{1973A&A....24..337S,1976MNRAS.175..613S}. Such an instability may be responsible for the outbursts observed in cataclysmic variables (CVs) and X-ray nova \citep*[e.g.,][]{1982A&A...106...34M,1982AcA....32..199S,1993adcs.book....6C}, which was also suggested as an possibility for accretion disk eruptions in AGNs \citep{1996ApJ...458..491S}. Furthermore, disk instability may also be the probably physical origin of the multiple changing-look AGNs challenged the AGN paradigm \citep{2018ApJ...862..109Y,2019ApJ...874....8M,2022arXiv221003928W}, though lots of model had been produced \citep{2015MNRAS.452...69M,2020ApJ...898L...1R,2020A&A...641A.167S,2020A&A...643L...9W,2021ApJ...910...97P,2022ApJ...927..227L}. The main difficulty of this disk instability model for QPEs is the viscous timescale of a thin accretion disk being significantly larger than the observed periods of QPEs (e.g., \citealt{2021ApJ...910...97P,2022ApJ...928L..18P}). It was suggested that the viscous timescale of a disk driven predominantly by the magnetic outflows can be substantially shortened \citep{2013ApJ...765..149C, 2014ApJ...786....6L, 2019ApJ...886...92L,2021ApJ...916...61F,2022arXiv221100704K, 2022ApJ...928L..18P,2022arXiv220410067S}. 
\cite{2022ApJ...928L..18P} constructed an instability model of the disk with magnetically driven outflows for the QPE GSN 069, and both of its light curve and phased-resolved X-ray spectra have been fitted by their model fairly well. In this work, we employ this model to the other QPEs based with the archived observational data such as 
the periods of bursts, and spectra, etc.



\section{Model}


Similar with our previous work \citep{2022ApJ...928L..18P}, we consider a thin accretion disk with winds driven by large-scale magnetic fields around a spinning super massive black hole (SMBH), where the general relativistic correction factors, the general form for the viscous torque and non-zero torque condition at innermost stable circular orbit (ISCO) are adopted to modify our basic equations. The steady outer thin disk can be described as:

\begin{equation}
    \frac{\mathrm{d}\dot{M}}{\mathrm{d}R}+4\pi R\dot{m}_{\rm w}=0,
    \label{continuity}
\end{equation}
\begin{equation}
    -\frac{1}{2\pi}\frac{\mathrm{d}(\dot{M}l_{\rm k})}{\mathrm{d}R}-\frac{\mathrm{d}}{\mathrm{d}R}(R^{2}\mathscr{B}\mathscr{C}^{-1/2}\mathscr{D}T_{r\phi})+T_{\rm m} R=0,
    \label{angularmom}
\end{equation}
\begin{equation}
    P_{\rm tot}=(1+\frac{1}{\beta_1})(P_{\rm gas}+P_{\rm rad}),
    \label{EoS}
\end{equation}
\begin{equation}
    -\frac{3}{2}\Omega_{\rm k}T_{r\phi}\frac{\mathscr{BD}}{\mathscr{C}}=\frac{8acT_{\rm c}^{4}}{3\tau}.
    \label{energy}
\end{equation}

For a thin disk with relatively high accretion rates, the inner disk region dominated by radiation pressure will produce limit-cycle bursts. In some specific parameter space, this unstable zone can be limited in a narrow annulus \citep{2020A&A...641A.167S,2021ApJ...910...97P,2022ApJ...928L..18P}. The evolution equation of surface density and central temperature of this narrow zone can be written as
\begin{equation}
\begin{split}
    &\left[u^t-\frac{C_{\rm H}H\left(1-\beta_{2}\right)}{\Sigma\left(1+\beta_{2}\right)}\right]\frac{\mathrm{d}\Sigma}{\mathrm{d}t}+\frac{C_{\rm H}H\left(4-3\beta_{2}\right)}{T\left(1+\beta_{2}\right)}\frac{\mathrm{d}T}{\mathrm{d}t}\\
    &-\frac{\dot{M}_{0}-\dot{M}-4\pi R\dot{m}_{\rm w}\Delta R}{2\pi R\Delta R}=0,
    \label{sur_den_evolution}
\end{split}
\end{equation}
\begin{equation}
\begin{aligned}
    \frac{\mathrm{d} T}{\mathrm{d}t}=&\frac{T(Q^+-Q^--Q_{\rm adv})(1+\frac{1}{\beta_1})(1+\beta_2)}{2PHu^{t}(28-22.5\beta_2-1.5\beta_2^2+\frac{12-9\beta_2}{\beta_1})}\\
    &+2\frac{T\mathrm{d}\Sigma}{\Sigma\mathrm{d}t}\frac{4-3\beta_2+\frac{2-\beta_2}{\beta_1}}{28-22.5\beta_2-1.5\beta_2^2+\frac{12-9\beta_2}{\beta_1}},
    \label{temp_evolution}
\end{aligned}
\end{equation}
respectively. The meanings of all above symbols are the same as those in \cite{2022ApJ...928L..18P}.

The temperature of the disk derived from the X-ray continuum spectra in quiescent state detected in other four QPEs, $kT_{\rm disk}[=11.5\left(M/10^8M_{\odot}\right)^{-1/4}\dot{m}^{1/4} (\rm eV)]$, are all higher than $50 \rm eV$ with black hole mass roughly ranged from $10^{5}$ to $10^{6} M_{\odot}$ \citep{2020A&A...636L...2G,2021ApJ...921L..40C,2022ApJ...930..122C}. As argued by \citet{2022ApJ...928L..18P}, the maximum effective temperature of a thin accretion disk surrounding a black hole ($M>2\times 10^5 M_{\odot}$) is hard to exceed 50 eV. 
 It was argued that the disk radiation is more complex than a sum of blackbody emission from the disk, as the electron scattering opacity is much greater than absorption opacity at the inner disk (see, e.g., \citealt{2012MNRAS.420.1848D}). Thus, a color correction factor for the effective temperature is required in the calculations of the emergent spectrum of the disk. A precise calculation of the disk spectrum should include the vertical structure of disk, with which the full radiative transfer equation is to be solved. This is beyond the scope of this work, instead, we adopt a diluted blackbody as a reasonable approximation \citep{1995ApJ...445..780S}:
\begin{equation}
    F_{\nu}^{\rm db}=\frac{1}{f_{\rm cor}^{4}}\pi B_{\nu}(f_{\rm cor}T_{\rm eff}),
    \label{dilutedbb}
\end{equation}
where $f_{\rm cor}$ and $B_{\nu}$ are the hardening factor and Planck function, respectively. A typical value of hardening factor, $f_{\rm cor}\sim 1.7$, is usually adopted for black hole binaries (BHBs). In principle, it may vary with temperature, which can be written as \citep{2006ApJ...647..525D,2012MNRAS.420.1848D}:
\begin{equation}
    f_{\rm cor}\sim\left(72/T_{\rm keV}\right)^{1/9}.
    \label{harden_fac}
\end{equation}
This equation is valid for the accretion disk in AGN when $T_{\rm max} > 10^{5}\rm K$. However, when the disk temperature is sufficiently low, the disk spectra will return to the disk blackbody as electron scattering no longer dominates the opacity. Therefore it is necessary to calculate the hardening factor at each radius because the temperature is much lower than $10^{5} \rm K$ for the outer accretion disk in AGN. We set a threshold of disk temperature $T_{\rm disk}=10^{5} \rm K$. When $T_{\rm disk}>10^{5}\rm K$, we adopt equation (\ref{harden_fac}) to calculate $f_{\rm cor}$, otherwise $f_{\rm cor}=1$ is chosen. The discontinuity value of $f_{\rm cor}$ near the threshold makes the spectrum not very smoothly, but it has little effect on the X-ray band we are concerned about.

\section{results}

We try to extend our model in \citet{2022ApJ...928L..18P} to all the QPEs in this work. $a_{*}=0.98$ and $f=0.9$ are always adopted for convenient. All other parameters adopted for each QPEs are shown in Table \ref{tab:pars}. The spectra used in this work are corrected by the instrumental effective area and galaxy absorption, which is same with our previous work. Since the hardening factor used in this study varies with radius, we present here only the hardening factor $f_{\rm cor,in}$ at the inner radius of the outer stable disk in Table \ref{tab:pars}, as it has the greatest impact on the soft X-ray spectrum.

\begin{table*}[htbp]
    \centering
    \caption{Detailed parameter of our model}
    \begin{tabular*}{0.9\textwidth}{ccccccccc}
    \hline
    \hline
        Source & $M (\times 10^{5}M_{\odot})$ & $\dot{m} (\dot{M}_{\rm Edd})$ & $\alpha$ & $\beta_{1}$ & $\mu$ & $n_{\rm H} (\times10^{20}\rm cm^{-2})$ & $\Delta R (R_{s})$ & $f_{\rm cor,in}$\\
    \hline
        RX J1301.9+2747 (2000) & 3 & 0.15 & 0.15 & 5.5 & 0.1 & 0.4 & 0.088 & 2.37 \\
        RX J1301.9+2747 (2019) & 3 & 0.15 & 0.15 & 5 & 0.1 & 0.4 & 0.083 & 2.41 \\
        eRO-QPE1 & 10 & 0.15 & 0.1 & 8 & 0.15 & 2 & 0.048 & 2.51 \\
        eRO-QPE2 & 1 & 0.19 & 0.1 & 10.5 & 0.1 & 10 & 0.119 & 2.24 \\
        XMMSL1 J024916.6-041244 & 0.7 & 0.08 & 0.15 & 24.5 & 0.22 & 6 & 0.099 & 2.19 \\
        GSN 069 & 6 & 0.11 & 0.15 & 4.5 & 0.13 & 1 & 0.047 & 2.5 \\
        GSN 069 (2) & 4 & 0.1 & 0.15 & 18 & 0.24 & 1 & 0.067 & 1.55\\
    \hline
    \end{tabular*}
    \label{tab:pars}
\end{table*}

\subsection{RX J1301.9+2747}

Several rapid flares have been observed in this source from 1991 to 2019 \citep{2000MNRAS.318..309D, 2013ApJ...768..167S, 2020A&A...636L...2G}. Compared with GSN 069, the flare recurrence time of RX J1301.9+2747 appears to be more complex and evolve rapidly. We suppose that its burst mechanism is the same with GSN 069 and that the evolution of light curve may be caused by the disturbance of accretion rate or magnetic field strength.

\begin{figure}[htbp]
   \centering
   \includegraphics[width=0.47\textwidth]{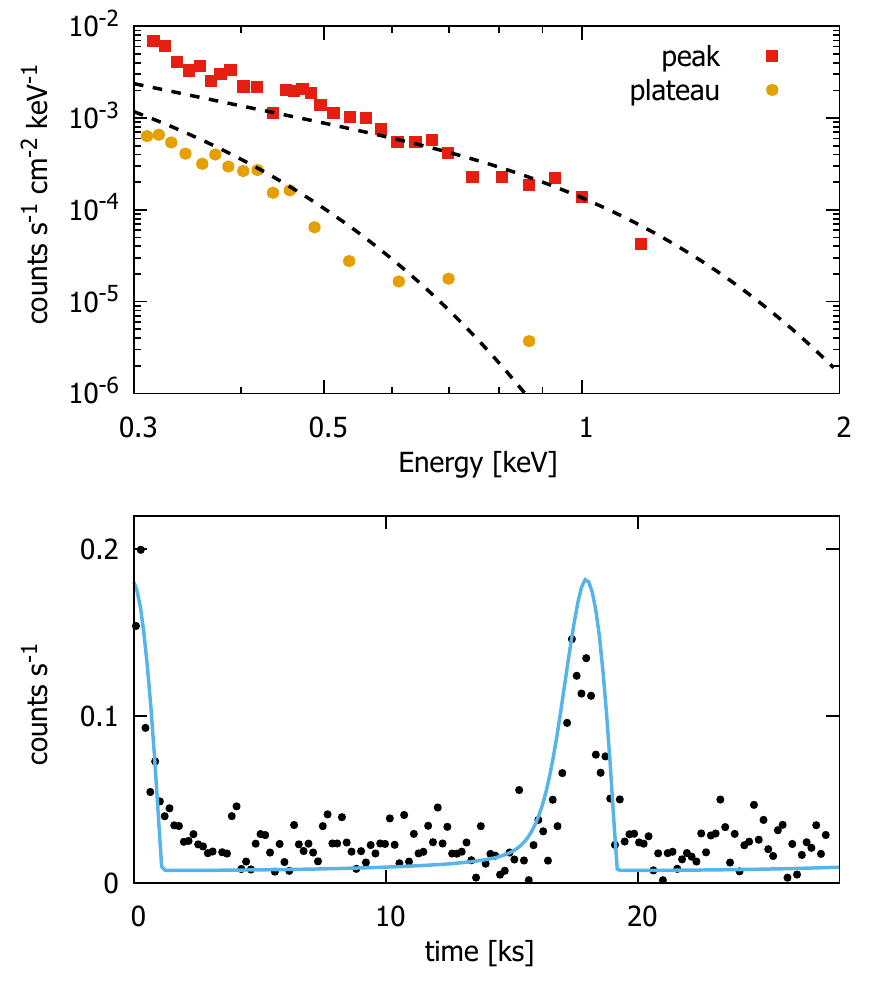}
   \caption{Phase-resolved Spectral analysis and $0.2-2 \rm keV$ light curve of RX J1301.9+2747 during the 10-11 December 2000. Upper panel: The red square and the yellow circles selected from eruptions represent the peak phase and the plateau phase, respectively. The dashed lines are given by our model. All the observational data are corrected by introducing instrumental effective area and absorption. Lower panel: The black dots and blue line represent the data obtained from XMM-Newton observations (EPIC-MOS1) and the light curve produced by our model, respectively.}
   \label{fig:rxj_2000}
\end{figure}

\begin{figure}[htbp]
   \centering
   \includegraphics[width=0.47\textwidth]{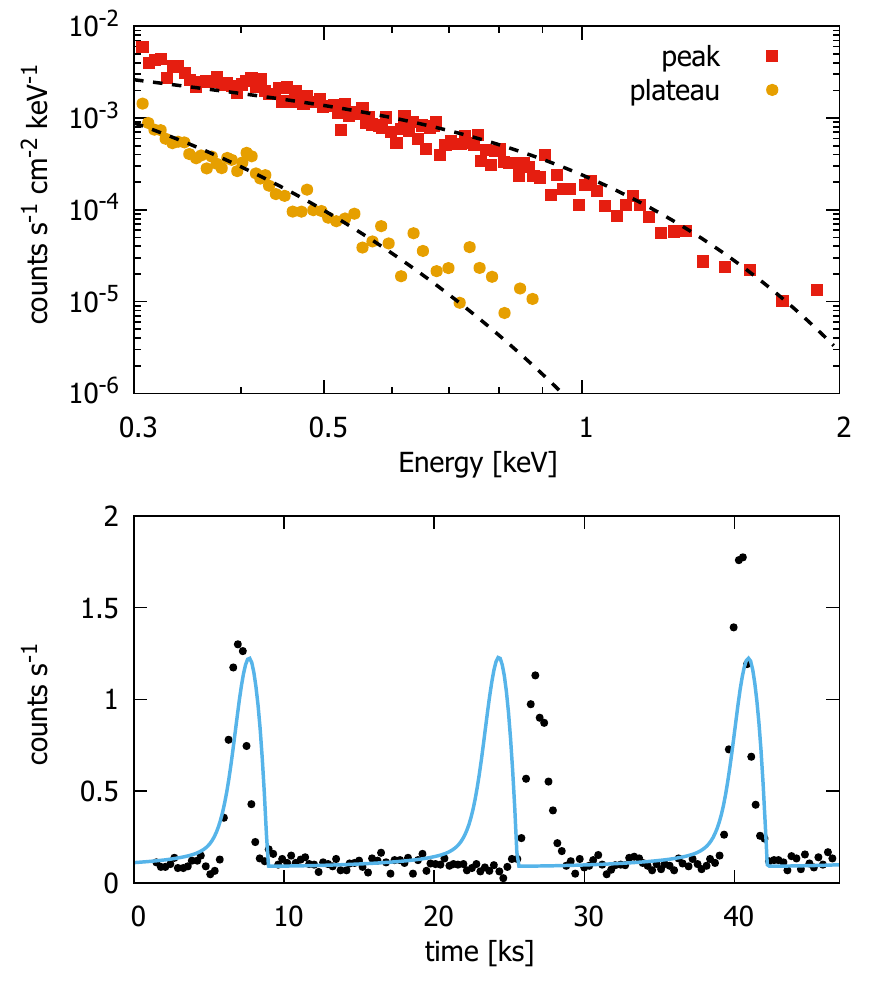}
   \caption{Same with Figure \ref{fig:rxj_2000}, but observed during the 30-31 May 2019.}
   \label{fig:rxj_2019}
\end{figure}

Two sets of parameters with the same black hole mass are adopted to fit the spectral evolution and light curve of RX J1301.9+2747 in 2000 and 2019, respectively. As shown in Figure \ref{fig:rxj_2000} and \ref{fig:rxj_2019}, our model can roughly reproduce the observational data. Note that the data of light curve in Figure \ref{fig:rxj_2000} is taken from EPIC-MOS1, since the EPIC-PN data only covered one eruption in 2000. Except for Figure \ref{fig:rxj_2000}, all other XMM-Newton data comes from EPIC-PN. The mass of this source was suggested as $8\times10^{5}M_{\odot}$ with a scatter of 0.5 dex \citep{2013ApJ...768..167S}, which is roughly consistent with the value $3\times10^{5}M_{\odot}$ in table 1. Since the two recurrence times in 2019 are distinct with each other (Figure \ref{fig:rxj_2019}), we take the average interval time as the period of our model to calculate the limit-cycle.

\subsection{eRO-QPE1}

\begin{figure}[htbp]
   \centering
   \includegraphics[width=0.47\textwidth]{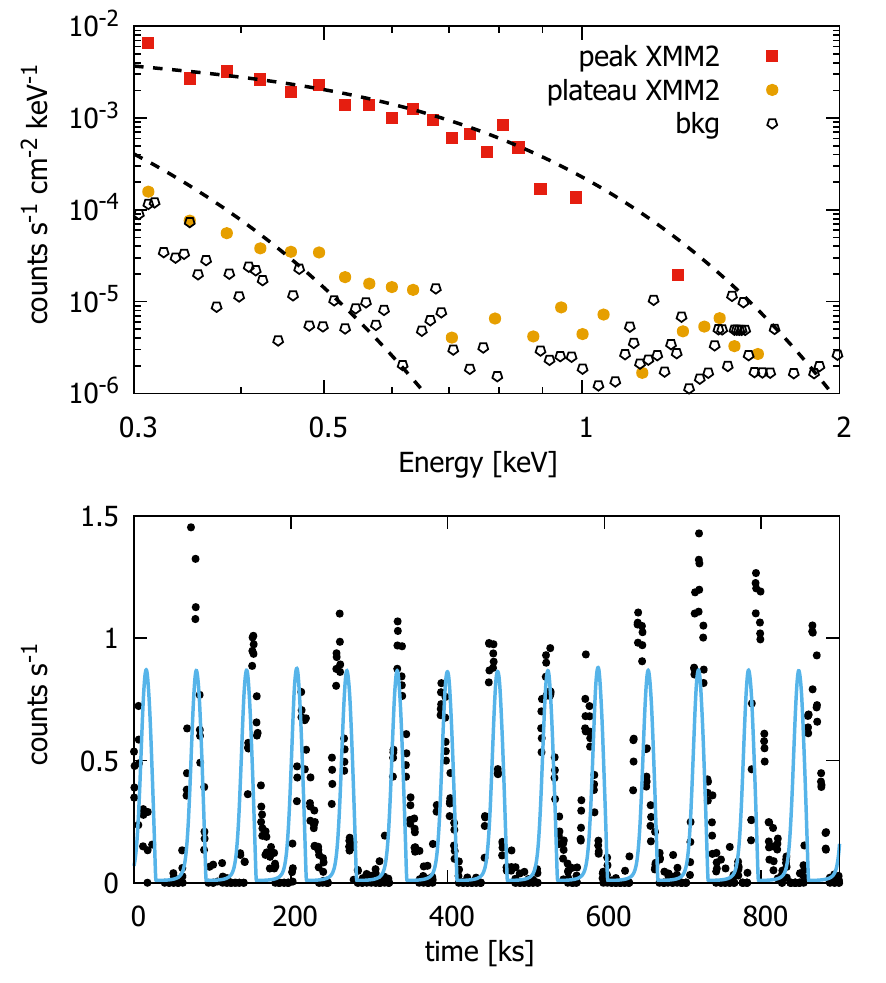}
   \caption{Phase-resolved Spectral analysis and $0.3-1 \rm keV$ light curve of eRO-QPE1. The spectral data was obtained from XMM-Newton on 4 August 2020 and the light curve data was observed by NICER on 19 August 2020.}
   \label{fig:QPE_1}
\end{figure}

This source had four observational campaigns, i.e., one of eROSITA, two of XMM-Newton and one of NICER, of which the quality of XMM-Newton data is the best \citep{2021Natur.592..704A}. Therefore we adopt the data from XMM-Newton to analyse the spectral evolution. However, due to the short duration of XMM-Newton campaign, the data from NICER (including 15 complete eruptions) are adopted to investigate the light curve (see Figure \ref{fig:QPE_1}). There are two observations available from XMM-Newton. The first one on 27 July 2020 (eRO-QPE1-XMM1) showed a complex profile that seems to be formed by the overlapping of several eruptions, which cannot be simplified explained by an instability model. We thus adopt the observation on 4 August 2020 (eRO-QPE1-XMM2) showing a single isolated burst \citep{2022A&A...662A..49A}, to compare with our spectral result.

In Figure \ref{fig:QPE_1}, we compare the calculated spectrum and light curves with the observations. The mass of black hole in this source is aopted as $M_{\rm BH}=1\times10^{6}M_{\odot}$, which is consistent with the estimation of \citet{2022ApJ...930..122C} ($\sim 9.1\times10^{5}M_{\odot}$) by using a empirical scaling relation. Excluding the effect of background, both the spectrum and light curves can qualitatively be reproduced by our model.


\subsection{eRO-QPE2}

\begin{figure}[htbp]
   \centering
   \includegraphics[width=0.47\textwidth]{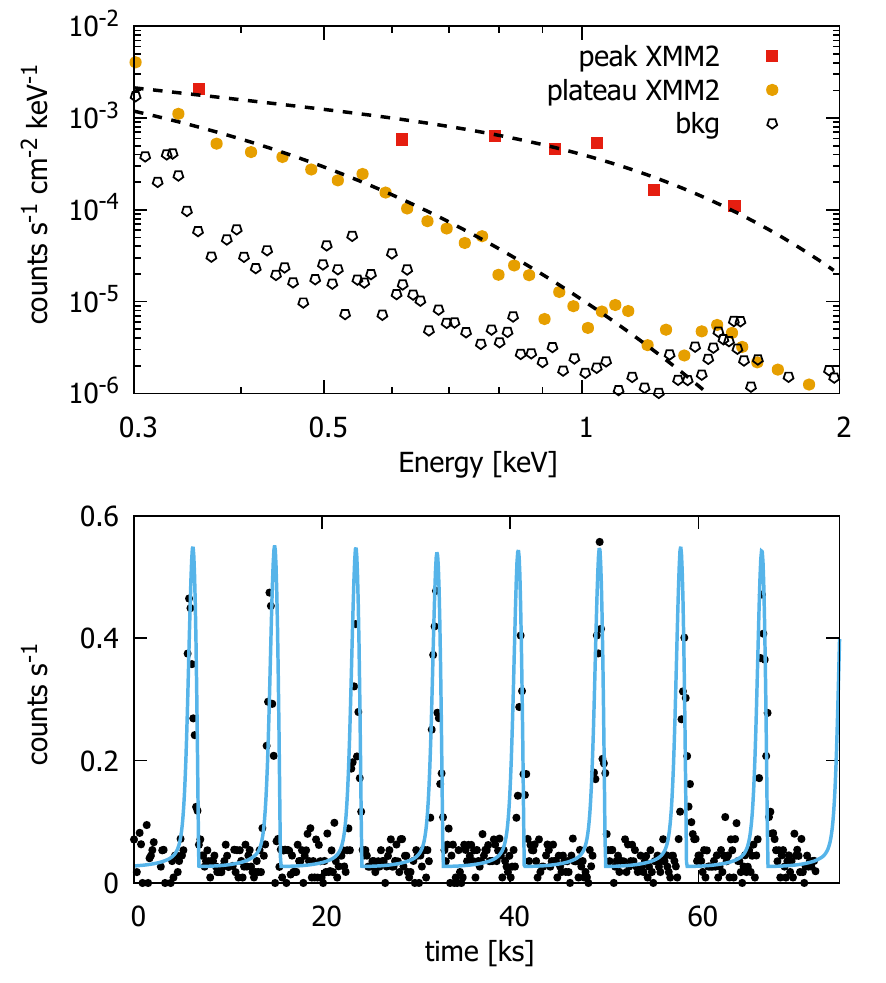}
   \caption{Phase-resolved Spectral analysis and $0.2-10 \rm keV$ light curve of eRO-QPE2. Both of spectral and light curve data was obtained from XMM-Newton on 6 August 2020.}
   \label{fig:QPE_2}
\end{figure}

The timing property and spectral evolution of this source seem to be similar with GSN 069, i.e., showing a small duty cycle and alternating longer and shorter recurrence times. But the period of eruptions in eRO-QPE2 is much shorter and the temperature of spectra is much higher than those in GSN 069. We can therefore infer that the mass of eRO-QPE2 should be smaller than GSN 069. In our calculation, a black hole mass $M_{\rm BH}=1\times10^{5}M_{\odot}$ is adopted, smaller than the mass of eRO-QPE2 adopted by \citet{2022ApJ...930..122C} ($2.3\times10^{5}M_{\rm BH}$).

We present the results in Figure \ref{fig:QPE_2}. Both the spectral evolution and light curves are well reproduced by our model, just like the results of our previous work for GSN 069.

\subsection{XMMSL1 J024916.6-041244}

This source is regarded as the most probable QPEs because it showed 1.5 QPE-like flares in 2006 and had a spectral evolution similar to that of GSN 069. 

\begin{figure}[htbp]
   \centering
   \includegraphics[width=0.47\textwidth]{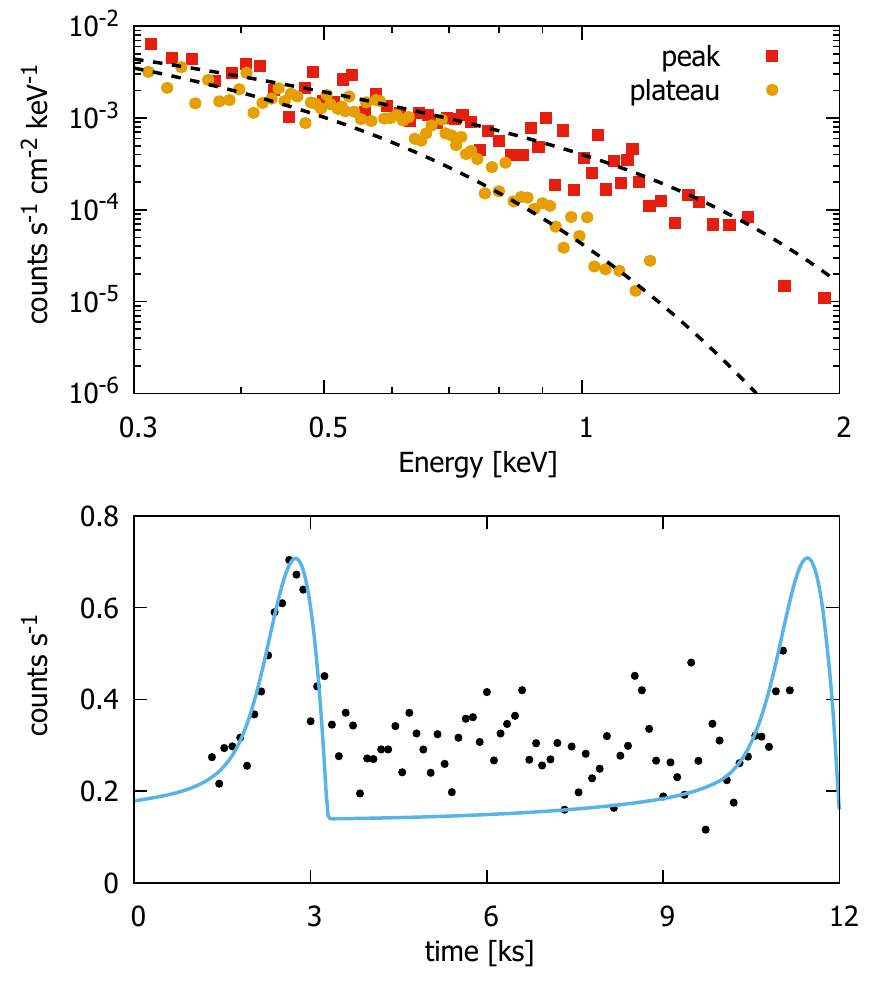}
   \caption{Phase-resolved Spectral analysis and $0.3-2 \rm keV$ light curve of XMMSL1 JJ024916.6-041244. Both of spectral and light curve data was obtained from XMM-Newton on 14 July 2006.}
   \label{fig:XMM}
\end{figure}

Two black hole mass of XMMSL1 J024916.6-041244 are given by the previous works \citep{2016A&A...592A..74S,2019MNRAS.487.4136W}, i.e.,  $M_{\rm BH}\sim8.5\times10^{4}M_{\odot}$ and $M_{\rm BH}\sim5\times10^{5}M_{\odot}$. However, the black hole mass in this source should be much smaller than that of GSN 069 due to its higher temperature in low state. Here we adopt $M_{\rm BH}=7\times10^{4}M_{\odot}$. Figure \ref{fig:XMM} gives the comparison of our model  with observations. We find that our model can qualitatively reproduce the light curve and spectral evolution as a whole.

\subsection{GSN 069}

\begin{figure}[htbp]
   \centering
   \includegraphics[width=0.47\textwidth]{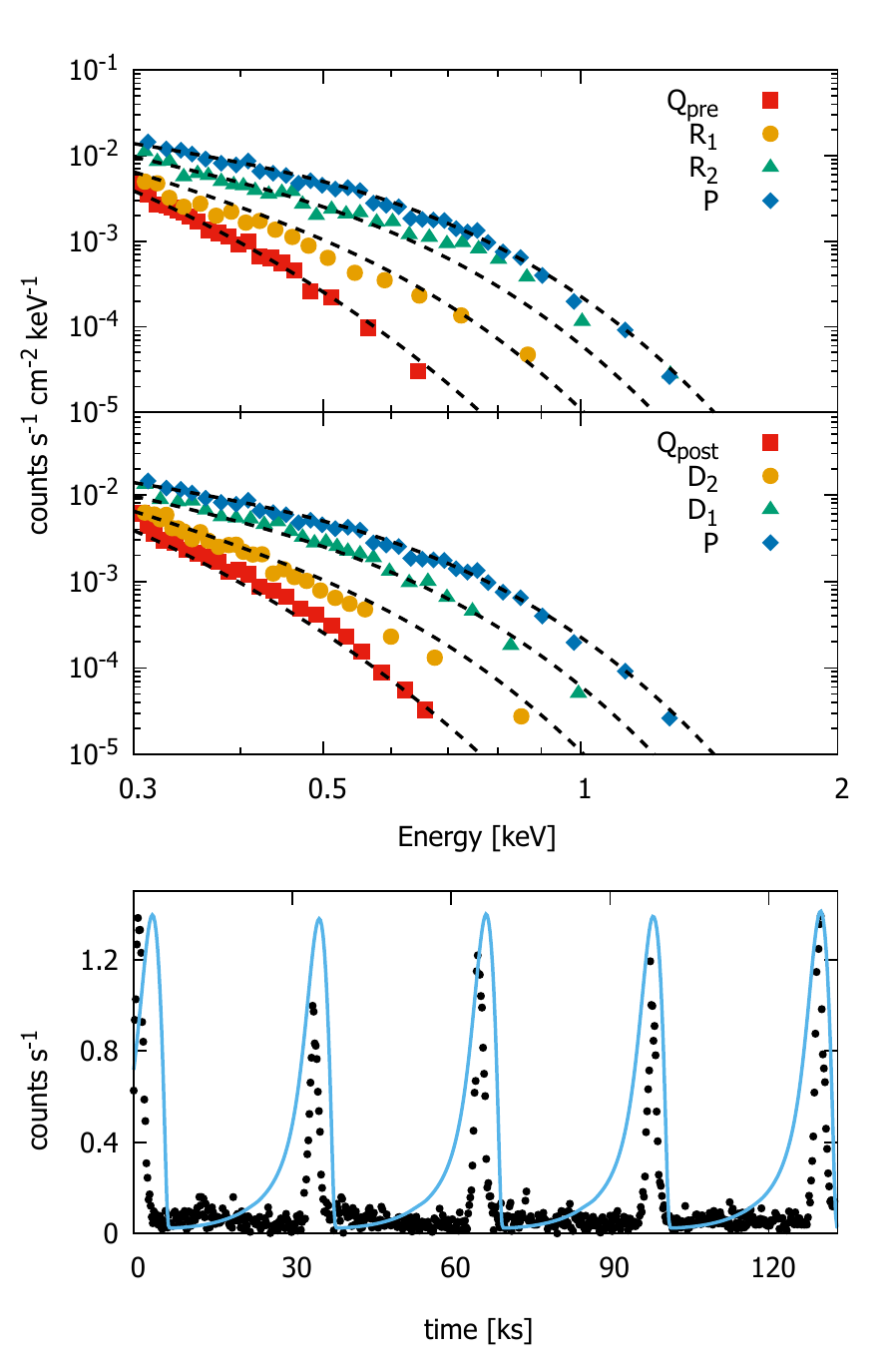}
   \caption{Phase-resolved Spectral analysis and $0.4-2 \rm keV$ light curve of GSN 069. The seven spectral segments are same as in \cite{2019Natur.573..381M}.}
   \label{fig:GSN}
\end{figure}

To investigate the effect of hardening factor on GNS 069, we also compare our numerical results with the observed light curve and X-ray spectra in Figure \ref{fig:GSN}. It is found that, by including the hardening factor, we can still achieve satisfactory results with a higher black hole mass ($M_{\rm BH}=6\times 10^{5}M_{\odot}$) compared with those in \citet{2022ApJ...928L..18P}.


\section{Conclusions and Discussion}

In this work, we adopt the model of \citet{2022ApJ...928L..18P} to fit the observed X-ray spectra of the five QPEs, while we improve the calculations the emergent spectra of the disk in this work. A hardening factor ($f_{\rm cor}$) is induced to modify the effective temperature of accretion disk based on the model of \citet{2022ApJ...928L..18P}, which is reasonable because the radiative spectrum of an accretion disk is much more complex than a simple sum of disk blackbody. Several methods had been proposed to estimate the value of hardening factor (e.g., \citealt{2002ApJ...572...79C,2006ApJ...647..525D,2012MNRAS.420.1848D,2019ApJ...874...23D}). We adopt the formula given by \citet{2012MNRAS.420.1848D} in this work rather than the more recent one in \citet{2019ApJ...874...23D} for the reason that our model involves a high spin black hole with inner disk temperature of $\sim 2\times10^{5}\rm K$, while the updated fitting equation provided by \citet{2019ApJ...874...23D} is designed for a non-spinning black hole and an accretion disk with higher temperature \citep{2022ApJ...939L...2Z}. Furthermore, it is found that the results of our model depend relatively weak on the value of hardening factor. For example, we also adopt the hardening factor given by equation (9) in \citet{2019ApJ...874...23D} to fit the observational data of GSN 069, where the parameters employed are shown in Table \ref{tab:pars} GSN 069 (2). A similar good fitting result can be achieved by slightly reducing the mass of black hole. Therefore, we adopt the color correction of \citet{2012MNRAS.420.1848D} for all of our other calculations. It is found that all the five QPEs can be qualitatively described by this model. However, there are still some inconsistencies in details, e.g., the profile and the irregular period of eruptions, which may be partly solved by considering more physical process in the model (see \citealt{2022ApJ...928L..18P} for details).


In general, the import of hardening factor can equivalently increase the effective temperature of disk and thus harden the radiation spectrum of the accretion disk \citep{2012MNRAS.420.1848D}, which can help to solve the inconsistency between the temperature given by a standard thin disk and that required by observations (see section 2 for details). The hardening factors adopted in this work are all within the range of 2.1 to 2.6, which is also consistent with previous works (e.g., \citealt{1992MNRAS.258..189R,2012MNRAS.420.1848D}). In addition, although the presence of large-scale magnetic fields can drive outflows from disk, the disk structure can always remain optically thick required to generate a diluted black-body spectrum. The reason is that the magnetic pressure is far smaller than the sum of gas and radiation pressure in our model. Indeed, the magnetically driven outflows in this case will reduce the temperature of accretion disk and simultaneously increase the surface density of disk in the inner region, resulting on the increase of effective optical depth (see \citealt{2014ApJ...788...71L} for details).

We need to emphasis that the spin and mass of black hole are coupled in some ways when fitting the observational data. A higher spin adopted in our model will produce a disk closer to the central black hole, implying a higher temperature in the inner disk and a shorter timescale of limit-cycle behaviour. This trend can be equivalently obtained by adopting a smaller black hole mass. However, a somewhat smaller black hole mass compared with that inferred from observations (but within the error bars) has to be usually adopted in order to accord with the effective temperature observed in low state, even we have adopted a high black hole spin.

\begin{figure}[bp]
   \centering
   \includegraphics[width=0.47\textwidth]{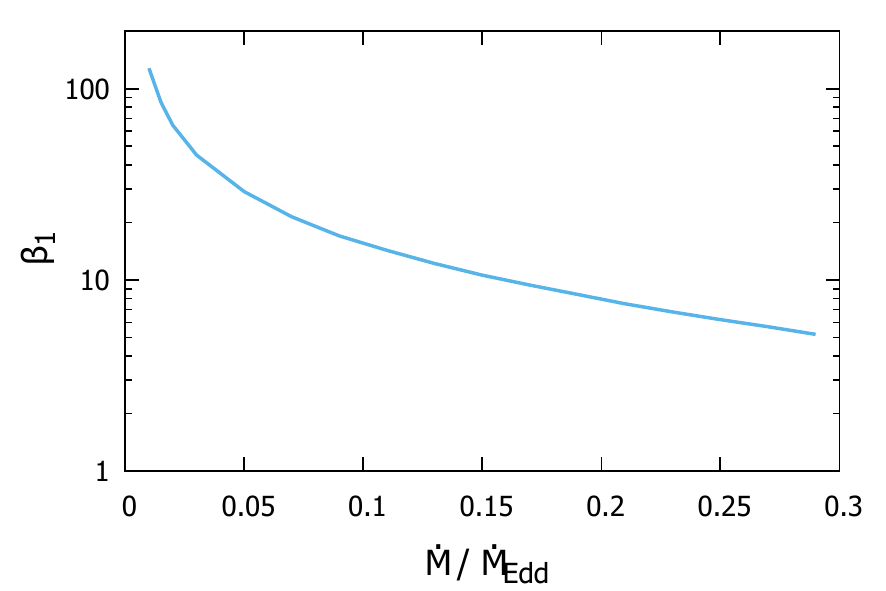}
   \caption{The distribution of magnetic field strength and mass accretion rate with a fixed unstable region width of $0.1R_{\rm s}$. Other parameters are fixed as: $M=6\times10^{5}M_{\odot}$, $\alpha=0.15$, $\mu=0.15$, $f=0.9$, $a_{*}=0.98$.}
   \label{fig:range}
\end{figure}

Two in the total five QPEs, i.e., GSN 069 and XMMSL1 J024916.6-041244, appear the long-term light curves consistent with TDEs, of which the fraction is much higher than the normal galaxy. Therefore, some authors suggested that QPEs may be triggered by the partial disruption of a remnant core or star after TDEs \citep{2022arXiv220707511M}. 
If the TDEs did happen in these two sources, the disk instability may still at work when the gas is being swallowed by the BHs. At the beginning of the decay phase in TDE, the accretion rate could be super-Eddington corresponding with the slim disk \citep{1988ApJ...332..646A}, which is thermally stable. However, the disk will slowly become the unstable thin accretion disk with decreasing accretion rate. QPEs will appear when the unstable region of disk is small enough with suitable accretion rate and magnetic fields. Analyzing the detailed parameter range that can generate QPE is very complicated because we have lots of parameters in this work. However, we can roughly constraint part of the parameters, such as mass accretion rate and magnetic pressure, by fixing other parameters. For example, we present how $\beta_{1}$ vary with mass accretion rate in GSN 069 by fixing the unstable region as $0.1R_{\rm s}$ and other parameters (see Figure \ref{fig:range}). However, we need to note that the period of QPE depends not only on the width of the unstable region, mass accretion rate and magnetic pressure, but also on other parameters, such as the black hole mass. For QPE with different black hole mass and periods, it is therefore necessary to adopt different parameters in order to achieve good fittings.


\cite{2022arXiv220707511M} showed that the QPEs of GSN 069 disappeared when a rebrightening of quiescent state happened after January 2020. According to the above discussion, the unstable zone in a thin disk will expand to larger radius, resulting on the increase of periods in QPEs when the accretion rate increases. Therefore, if the rebrightening of GSN 069 is caused by a repeating TDEs, the period of QPEs will be greatly increased with the sudden increase of accretion rate. So the QPEs can hardly be observed in a short duration campaign. The problem is that the period remains almost constant while the amplitude of QPEs in the XMM6 decays linearly with time at the beginning of rebrightening \citep{2022arXiv220707511M}, which is inconsistent with the disk instability model. We suggest that the flux at the fast rising phase of rebrightening may come from the stream shocks at apocentre by TDEs, the disk instability model works only when the mass accretion rate has decreased to a critical value.


\section* {ACKNOWLEDGEMENTS}
We thank the reviewer for helpful comments. XP thanks Dr. Riccardo Arcodia, Erin Kara, Joheen Chakraborty and Margherita Giustini for providing of data in our figures. This work is supported by the NSFC (grants 12273089, 12073023, 12233007, 11833007, and 12147103), the science research grants from the China Manned Space Project with No. CMS-CSST- 2021-A06, and the Fundamental Research Fund for Chinese Central Universities. 

\bibliographystyle{aasjournal}
\bibliography{ref}{}

\begin{thebibliography}{}
\expandafter\ifx\csname natexlab\endcsname\relax\def\natexlab#1{#1}\fi
\providecommand{\url}[1]{\href{#1}{#1}}
\providecommand{\dodoi}[1]{doi:~\href{http://doi.org/#1}{\nolinkurl{#1}}}
\providecommand{\doeprint}[1]{\href{http://ascl.net/#1}{\nolinkurl{http://ascl.net/#1}}}
\providecommand{\doarXiv}[1]{\href{https://arxiv.org/abs/#1}{\nolinkurl{https://arxiv.org/abs/#1}}}

\bibitem[{{Abramowicz} {et~al.}(1988){Abramowicz}, {Czerny}, {Lasota}, \&
  {Szuszkiewicz}}]{1988ApJ...332..646A}
{Abramowicz}, M.~A., {Czerny}, B., {Lasota}, J.~P., \& {Szuszkiewicz}, E. 1988,
  \apj, 332, 646, \dodoi{10.1086/166683}

\bibitem[{{Arcodia} {et~al.}(2021){Arcodia}, {Merloni}, {Nandra}, {Buchner},
  {Salvato}, {Pasham}, {Remillard}, {Comparat}, {Lamer}, {Ponti}, {Malyali},
  {Wolf}, {Arzoumanian}, {Bogensberger}, {Buckley}, {Gendreau}, {Gromadzki},
  {Kara}, {Krumpe}, {Markwardt}, {Ramos-Ceja}, {Rau}, {Schramm}, \&
  {Schwope}}]{2021Natur.592..704A}
{Arcodia}, R., {Merloni}, A., {Nandra}, K., {et~al.} 2021, \nat, 592, 704,
  \dodoi{10.1038/s41586-021-03394-6}

\bibitem[{{Arcodia} {et~al.}(2022){Arcodia}, {Miniutti}, {Ponti}, {Buchner},
  {Giustini}, {Merloni}, {Nandra}, {Vincentelli}, {Kara}, {Salvato}, \&
  {Pasham}}]{2022A&A...662A..49A}
{Arcodia}, R., {Miniutti}, G., {Ponti}, G., {et~al.} 2022, \aap, 662, A49,
  \dodoi{10.1051/0004-6361/202243259}

\bibitem[{{Bonnerot} \& {Stone}(2021)}]{2021SSRv..217...16B}
{Bonnerot}, C., \& {Stone}, N.~C. 2021, \ssr, 217, 16,
  \dodoi{10.1007/s11214-020-00789-1}

\bibitem[{{Cannizzo}(1993)}]{1993adcs.book....6C}
{Cannizzo}, J.~K. 1993, in Accretion Disks in Compact Stellar Systems, 6--40,
  \dodoi{10.1142/9789814350976_0002}

\bibitem[{{Cao} \& {Spruit}(2013)}]{2013ApJ...765..149C}
{Cao}, X., \& {Spruit}, H.~C. 2013, \apj, 765, 149,
  \dodoi{10.1088/0004-637X/765/2/149}

\bibitem[{{Chakraborty} {et~al.}(2021){Chakraborty}, {Kara}, {Masterson},
  {Giustini}, {Miniutti}, \& {Saxton}}]{2021ApJ...921L..40C}
{Chakraborty}, J., {Kara}, E., {Masterson}, M., {et~al.} 2021, \apjl, 921, L40,
  \dodoi{10.3847/2041-8213/ac313b}

\bibitem[{{Chen} {et~al.}(2022{\natexlab{a}}){Chen}, {Shen}, \&
  {Liu}}]{2022arXiv221009945C}
{Chen}, J.-H., {Shen}, R.-F., \& {Liu}, S.-F. 2022{\natexlab{a}}, arXiv
  e-prints, arXiv:2210.09945.
\newblock \doarXiv{2210.09945}

\bibitem[{{Chen} {et~al.}(2022{\natexlab{b}}){Chen}, {Qiu}, {Li}, \&
  {Liu}}]{2022ApJ...930..122C}
{Chen}, X., {Qiu}, Y., {Li}, S., \& {Liu}, F.~K. 2022{\natexlab{b}}, \apj, 930,
  122, \dodoi{10.3847/1538-4357/ac63bf}

\bibitem[{{Chiang}(2002)}]{2002ApJ...572...79C}
{Chiang}, J. 2002, \apj, 572, 79, \dodoi{10.1086/340193}

\bibitem[{{Davis} {et~al.}(2006){Davis}, {Done}, \&
  {Blaes}}]{2006ApJ...647..525D}
{Davis}, S.~W., {Done}, C., \& {Blaes}, O.~M. 2006, \apj, 647, 525,
  \dodoi{10.1086/505386}

\bibitem[{{Davis} \& {El-Abd}(2019)}]{2019ApJ...874...23D}
{Davis}, S.~W., \& {El-Abd}, S. 2019, \apj, 874, 23,
  \dodoi{10.3847/1538-4357/ab05c5}

\bibitem[{{Dewangan} {et~al.}(2000){Dewangan}, {Singh}, {Mayya}, \&
  {Anupama}}]{2000MNRAS.318..309D}
{Dewangan}, G.~C., {Singh}, K.~P., {Mayya}, Y.~D., \& {Anupama}, G.~C. 2000,
  \mnras, 318, 309, \dodoi{10.1046/j.1365-8711.2000.03755.x}

\bibitem[{{Done} {et~al.}(2012){Done}, {Davis}, {Jin}, {Blaes}, \&
  {Ward}}]{2012MNRAS.420.1848D}
{Done}, C., {Davis}, S.~W., {Jin}, C., {Blaes}, O., \& {Ward}, M. 2012, \mnras,
  420, 1848, \dodoi{10.1111/j.1365-2966.2011.19779.x}

\bibitem[{{Esquej} {et~al.}(2007){Esquej}, {Saxton}, {Freyberg}, {Read},
  {Altieri}, {Sanchez-Portal}, \& {Hasinger}}]{2007A&A...462L..49E}
{Esquej}, P., {Saxton}, R.~D., {Freyberg}, M.~J., {et~al.} 2007, \aap, 462,
  L49, \dodoi{10.1051/0004-6361:20066072}

\bibitem[{{Feng} {et~al.}(2021){Feng}, {Cao}, {Li}, \&
  {Gu}}]{2021ApJ...916...61F}
{Feng}, J., {Cao}, X., {Li}, J.-w., \& {Gu}, W.-M. 2021, \apj, 916, 61,
  \dodoi{10.3847/1538-4357/ac07a6}

\bibitem[{{Giustini} {et~al.}(2020){Giustini}, {Miniutti}, \&
  {Saxton}}]{2020A&A...636L...2G}
{Giustini}, M., {Miniutti}, G., \& {Saxton}, R.~D. 2020, \aap, 636, L2,
  \dodoi{10.1051/0004-6361/202037610}

\bibitem[{{Ingram} {et~al.}(2021){Ingram}, {Motta}, {Aigrain}, \&
  {Karastergiou}}]{2021MNRAS.503.1703I}
{Ingram}, A., {Motta}, S.~E., {Aigrain}, S., \& {Karastergiou}, A. 2021,
  \mnras, 503, 1703, \dodoi{10.1093/mnras/stab609}

\bibitem[{{Kaur} {et~al.}(2022){Kaur}, {Stone}, \&
  {Gilbaum}}]{2022arXiv221100704K}
{Kaur}, K., {Stone}, N.~C., \& {Gilbaum}, S. 2022, arXiv e-prints,
  arXiv:2211.00704.
\newblock \doarXiv{2211.00704}

\bibitem[{{King}(2020)}]{2020MNRAS.493L.120K}
{King}, A. 2020, \mnras, 493, L120, \dodoi{10.1093/mnrasl/slaa020}

\bibitem[{{King}(2022)}]{2022MNRAS.515.4344K}
---. 2022, \mnras, 515, 4344, \dodoi{10.1093/mnras/stac1641}

\bibitem[{{Krolik} \& {Linial}(2022)}]{2022arXiv220902786K}
{Krolik}, J.~H., \& {Linial}, I. 2022, arXiv e-prints, arXiv:2209.02786.
\newblock \doarXiv{2209.02786}

\bibitem[{{Li} \& {Cao}(2019)}]{2019ApJ...886...92L}
{Li}, J., \& {Cao}, X. 2019, \apj, 886, 92, \dodoi{10.3847/1538-4357/ab4c36}

\bibitem[{{Li}(2014)}]{2014ApJ...788...71L}
{Li}, S.-L. 2014, \apj, 788, 71, \dodoi{10.1088/0004-637X/788/1/71}

\bibitem[{{Li} \& {Begelman}(2014)}]{2014ApJ...786....6L}
{Li}, S.-L., \& {Begelman}, M.~C. 2014, \apj, 786, 6,
  \dodoi{10.1088/0004-637X/786/1/6}

\bibitem[{{Linial} \& {Sari}(2022)}]{2022arXiv221109851L}
{Linial}, I., \& {Sari}, R. 2022, arXiv e-prints, arXiv:2211.09851.
\newblock \doarXiv{2211.09851}

\bibitem[{{Lu} \& {Quataert}(2022)}]{2022arXiv221008023L}
{Lu}, W., \& {Quataert}, E. 2022, arXiv e-prints, arXiv:2210.08023.
\newblock \doarXiv{2210.08023}

\bibitem[{{Lyu} {et~al.}(2022){Lyu}, {Wu}, {Yan}, {Yu}, \&
  {Liu}}]{2022ApJ...927..227L}
{Lyu}, B., {Wu}, Q., {Yan}, Z., {Yu}, W., \& {Liu}, H. 2022, \apj, 927, 227,
  \dodoi{10.3847/1538-4357/ac5256}

\bibitem[{{MacLeod} {et~al.}(2019){MacLeod}, {Green}, {Anderson}, {Bruce},
  {Eracleous}, {Graham}, {Homan}, {Lawrence}, {LeBleu}, {Ross}, {Ruan},
  {Runnoe}, {Stern}, {Burgett}, {Chambers}, {Kaiser}, {Magnier}, \&
  {Metcalfe}}]{2019ApJ...874....8M}
{MacLeod}, C.~L., {Green}, P.~J., {Anderson}, S.~F., {et~al.} 2019, \apj, 874,
  8, \dodoi{10.3847/1538-4357/ab05e2}

\bibitem[{{Merloni} {et~al.}(2015){Merloni}, {Dwelly}, {Salvato},
  {Georgakakis}, {Greiner}, {Krumpe}, {Nandra}, {Ponti}, \&
  {Rau}}]{2015MNRAS.452...69M}
{Merloni}, A., {Dwelly}, T., {Salvato}, M., {et~al.} 2015, \mnras, 452, 69,
  \dodoi{10.1093/mnras/stv1095}

\bibitem[{{Metzger} \& {Stone}(2016)}]{2016MNRAS.461..948M}
{Metzger}, B.~D., \& {Stone}, N.~C. 2016, \mnras, 461, 948,
  \dodoi{10.1093/mnras/stw1394}

\bibitem[{{Metzger} {et~al.}(2022){Metzger}, {Stone}, \&
  {Gilbaum}}]{2022ApJ...926..101M}
{Metzger}, B.~D., {Stone}, N.~C., \& {Gilbaum}, S. 2022, \apj, 926, 101,
  \dodoi{10.3847/1538-4357/ac3ee1}

\bibitem[{{Meyer} \& {Meyer-Hofmeister}(1982)}]{1982A&A...106...34M}
{Meyer}, F., \& {Meyer-Hofmeister}, E. 1982, \aap, 106, 34

\bibitem[{{Miniutti} {et~al.}(2022){Miniutti}, {Giustini}, {Arcodia}, {Saxton},
  {Read}, {Bianchi}, \& {Alexander}}]{2022arXiv220707511M}
{Miniutti}, G., {Giustini}, M., {Arcodia}, R., {et~al.} 2022, arXiv e-prints,
  arXiv:2207.07511.
\newblock \doarXiv{2207.07511}

\bibitem[{{Miniutti} {et~al.}(2023){Miniutti}, {Giustini}, {Arcodia}, {Saxton},
  {Read}, {Bianchi}, \& {Alexander}}]{2023A&A...670A..93M}
---. 2023, \aap, 670, A93, \dodoi{10.1051/0004-6361/202244512}

\bibitem[{{Miniutti} {et~al.}(2019){Miniutti}, {Saxton}, {Giustini},
  {Alexander}, {Fender}, {Heywood}, {Monageng}, {Coriat}, {Tzioumis}, {Read},
  {Knigge}, {Gandhi}, {Pretorius}, \&
  {Ag{\'\i}s-Gonz{\'a}lez}}]{2019Natur.573..381M}
{Miniutti}, G., {Saxton}, R.~D., {Giustini}, M., {et~al.} 2019, \nat, 573, 381,
  \dodoi{10.1038/s41586-019-1556-x}

\bibitem[{{Pan} {et~al.}(2021){Pan}, {Li}, \& {Cao}}]{2021ApJ...910...97P}
{Pan}, X., {Li}, S.-L., \& {Cao}, X. 2021, \apj, 910, 97,
  \dodoi{10.3847/1538-4357/abe766}

\bibitem[{{Pan} {et~al.}(2022){Pan}, {Li}, {Cao}, {Miniutti}, \&
  {Gu}}]{2022ApJ...928L..18P}
{Pan}, X., {Li}, S.-L., {Cao}, X., {Miniutti}, G., \& {Gu}, M. 2022, \apjl,
  928, L18, \dodoi{10.3847/2041-8213/ac5faf}

\bibitem[{{Piran} {et~al.}(2015){Piran}, {Svirski}, {Krolik}, {Cheng}, \&
  {Shiokawa}}]{2015ApJ...806..164P}
{Piran}, T., {Svirski}, G., {Krolik}, J., {Cheng}, R.~M., \& {Shiokawa}, H.
  2015, \apj, 806, 164, \dodoi{10.1088/0004-637X/806/2/164}

\bibitem[{{Raj} \& {Nixon}(2021)}]{2021ApJ...909...82R}
{Raj}, A., \& {Nixon}, C.~J. 2021, \apj, 909, 82,
  \dodoi{10.3847/1538-4357/abdc25}

\bibitem[{{Ricci} {et~al.}(2020){Ricci}, {Kara}, {Loewenstein}, {Trakhtenbrot},
  {Arcavi}, {Remillard}, {Fabian}, {Gendreau}, {Arzoumanian}, {Li}, {Ho},
  {MacLeod}, {Cackett}, {Altamirano}, {Gandhi}, {Kosec}, {Pasham}, {Steiner},
  \& {Chan}}]{2020ApJ...898L...1R}
{Ricci}, C., {Kara}, E., {Loewenstein}, M., {et~al.} 2020, \apjl, 898, L1,
  \dodoi{10.3847/2041-8213/ab91a1}

\bibitem[{{Ross} {et~al.}(1992){Ross}, {Fabian}, \&
  {Mineshige}}]{1992MNRAS.258..189R}
{Ross}, R.~R., {Fabian}, A.~C., \& {Mineshige}, S. 1992, \mnras, 258, 189,
  \dodoi{10.1093/mnras/258.1.189}

\bibitem[{{Shakura} \& {Sunyaev}(1973)}]{1973A&A....24..337S}
{Shakura}, N.~I., \& {Sunyaev}, R.~A. 1973, \aap, 24, 337

\bibitem[{{Shakura} \& {Sunyaev}(1976)}]{1976MNRAS.175..613S}
---. 1976, \mnras, 175, 613, \dodoi{10.1093/mnras/175.3.613}

\bibitem[{{Sheng} {et~al.}(2021){Sheng}, {Wang}, {Ferland}, {Shu}, {Yang},
  {Jiang}, \& {Chen}}]{2021ApJ...920L..25S}
{Sheng}, Z., {Wang}, T., {Ferland}, G., {et~al.} 2021, \apjl, 920, L25,
  \dodoi{10.3847/2041-8213/ac2251}

\bibitem[{{Shimura} \& {Takahara}(1995)}]{1995ApJ...445..780S}
{Shimura}, T., \& {Takahara}, F. 1995, \apj, 445, 780, \dodoi{10.1086/175740}

\bibitem[{{Shu} {et~al.}(2018){Shu}, {Wang}, {Dou}, {Jiang}, {Wang}, \&
  {Wang}}]{2018ApJ...857L..16S}
{Shu}, X.~W., {Wang}, S.~S., {Dou}, L.~M., {et~al.} 2018, \apjl, 857, L16,
  \dodoi{10.3847/2041-8213/aaba17}

\bibitem[{{Siemiginowska} {et~al.}(1996){Siemiginowska}, {Czerny}, \&
  {Kostyunin}}]{1996ApJ...458..491S}
{Siemiginowska}, A., {Czerny}, B., \& {Kostyunin}, V. 1996, \apj, 458, 491,
  \dodoi{10.1086/176831}

\bibitem[{{Smak}(1982)}]{1982AcA....32..199S}
{Smak}, J. 1982, \actaa, 32, 199

\bibitem[{{Sniegowska} {et~al.}(2020){Sniegowska}, {Czerny}, {Bon}, \&
  {Bon}}]{2020A&A...641A.167S}
{Sniegowska}, M., {Czerny}, B., {Bon}, E., \& {Bon}, N. 2020, \aap, 641, A167,
  \dodoi{10.1051/0004-6361/202038575}

\bibitem[{{{\'S}niegowska} {et~al.}(2022){{\'S}niegowska},
  {Grz{\k{e}}dzielski}, {Czerny}, \& {Janiuk}}]{2022arXiv220410067S}
{{\'S}niegowska}, M., {Grz{\k{e}}dzielski}, M., {Czerny}, B., \& {Janiuk}, A.
  2022, arXiv e-prints, arXiv:2204.10067.
\newblock \doarXiv{2204.10067}

\bibitem[{{Strotjohann} {et~al.}(2016){Strotjohann}, {Saxton}, {Starling},
  {Esquej}, {Read}, {Evans}, \& {Miniutti}}]{2016A&A...592A..74S}
{Strotjohann}, N.~L., {Saxton}, R.~D., {Starling}, R.~L.~C., {et~al.} 2016,
  \aap, 592, A74, \dodoi{10.1051/0004-6361/201628241}

\bibitem[{{Sukov{\'a}} {et~al.}(2021){Sukov{\'a}}, {Zaja{\v{c}}ek}, {Witzany},
  \& {Karas}}]{2021ApJ...917...43S}
{Sukov{\'a}}, P., {Zaja{\v{c}}ek}, M., {Witzany}, V., \& {Karas}, V. 2021,
  \apj, 917, 43, \dodoi{10.3847/1538-4357/ac05c6}

\bibitem[{{Sun} {et~al.}(2013){Sun}, {Shu}, \& {Wang}}]{2013ApJ...768..167S}
{Sun}, L., {Shu}, X., \& {Wang}, T. 2013, \apj, 768, 167,
  \dodoi{10.1088/0004-637X/768/2/167}

\bibitem[{{Wang} {et~al.}(2022{\natexlab{a}}){Wang}, {Xu}, {Bai}, {Brink},
  {Gao}, {Zheng}, \& {Filippenko}}]{2022arXiv221003928W}
{Wang}, J., {Xu}, D.~W., {Bai}, J.~Y., {et~al.} 2022{\natexlab{a}}, arXiv
  e-prints, arXiv:2210.03928.
\newblock \doarXiv{2210.03928}

\bibitem[{{Wang} \& {Bon}(2020)}]{2020A&A...643L...9W}
{Wang}, J.-M., \& {Bon}, E. 2020, \aap, 643, L9,
  \dodoi{10.1051/0004-6361/202039368}

\bibitem[{{Wang} {et~al.}(2022{\natexlab{b}}){Wang}, {Yin}, {Ma}, \&
  {Wu}}]{2022ApJ...933..225W}
{Wang}, M., {Yin}, J., {Ma}, Y., \& {Wu}, Q. 2022{\natexlab{b}}, \apj, 933,
  225, \dodoi{10.3847/1538-4357/ac75e6}

\bibitem[{{Wevers} {et~al.}(2019){Wevers}, {Stone}, {van Velzen}, {Jonker},
  {Hung}, {Auchettl}, {Gezari}, {Onori}, {Mata S{\'a}nchez},
  {Kostrzewa-Rutkowska}, \& {Casares}}]{2019MNRAS.487.4136W}
{Wevers}, T., {Stone}, N.~C., {van Velzen}, S., {et~al.} 2019, \mnras, 487,
  4136, \dodoi{10.1093/mnras/stz1602}

\bibitem[{{Xian} {et~al.}(2021){Xian}, {Zhang}, {Dou}, {He}, \&
  {Shu}}]{2021ApJ...921L..32X}
{Xian}, J., {Zhang}, F., {Dou}, L., {He}, J., \& {Shu}, X. 2021, \apjl, 921,
  L32, \dodoi{10.3847/2041-8213/ac31aa}

\bibitem[{{Yang} {et~al.}(2018){Yang}, {Wu}, {Fan}, {Jiang}, {McGreer},
  {Shangguan}, {Yao}, {Wang}, {Joshi}, {Green}, {Wang}, {Feng}, {Fu}, {Yang},
  \& {Liu}}]{2018ApJ...862..109Y}
{Yang}, Q., {Wu}, X.-B., {Fan}, X., {et~al.} 2018, \apj, 862, 109,
  \dodoi{10.3847/1538-4357/aaca3a}

\bibitem[{{Zdziarski} {et~al.}(2022){Zdziarski}, {You}, \&
  {Szanecki}}]{2022ApJ...939L...2Z}
{Zdziarski}, A.~A., {You}, B., \& {Szanecki}, M. 2022, \apjl, 939, L2,
  \dodoi{10.3847/2041-8213/ac9474}

\end{thebibliography}



\end{document}